# Reliable Density Functional Theory Predictions of Bandgaps for Materials

Chenxi Lu[1], Musen Li[2,3], Michael J. Ford,[2] Rika Kobayashi[4], Roger D. Amos[5], and Jeffrey R. Reimers[1,2]

1 International Centre for Quantum and Molecular Structures and Department of Physics, Shanghai University, Shanghai 200444, China.
2 University of Technology Sydney, School of Mathematical and Physical Sciences, Ultimo, New South Wales 2007, Australia.
3 Materials Genome Institute, Shanghai University, Shanghai 200444, China.
[4] ANU Supercomputer Facility, Leonard Huxley Bldg. 56, Mills Rd, Canberra, ACT, 2601. Australia.
[5] University of New South Wales (Canberra), Canberra, ACT, 2601
Email: jeffrey.reimers@uts.edu.au, Rika.Kobayashi@anu.edu.au, R.Amos@unsw.edu.au

**ABSTRACT** We consider methods for optimizing the bandgap calculation of 3D materials, considering 340 sample materials. Examined are the effects of the choice of the pseudopotential to describe core electrons, the plane-wave basis set cutoff energy, and the Brillouin zone integration. Cost-saving calculations in which the structure is optimized using reduced-quality Brillouin zone integrations and cutoff energies were found to lead to experimentally significant errors exceeding 0.1 eV in 18% of cases using the PBE functional and 21% of cases using PBE0. Such cost-savings approaches are therefore not recommended for general applications. Also, the current practice of using unoptimized grids to perform the Brillouin-zone integrations in bandgap calculations is found to be unreliable for 16% of materials using PBE and for 23% using PBE0. A **k**-space optimization scheme is introduced that interpolates extensive PBE results to determine a generally useful approach that when used in PBE0 calculations is found to be inadequate for only 1.6% of the materials studied.

# 1. INTRODUCTION

Three dimensional materials have attracted extensive research owing to their interesting photon properties [1], and have extremely broad application prospects in fields such as displays [1] and energy-storage batteries [2]. A material's bandgap influences both the highest wavelength at which it absorbs light and the energy that it can store. This then affects properties such as the efficiency of solar cells [3] and the transmission performance of optical fibers [4] with indeed materials with specific bandgaps being required in solar cells [5] and fiber optic sensors [6]. For semiconductors, a material's bandgap also influences its conductivity and usefulness [7]. In general, choosing a specific bandgap material directly affects the performance of the semiconductor device [8]. The reliable prediction of bandgaps is therefore a computation challenge of great value.

The two most widely used approaches to predicting bandgaps for materials are density functional theory [9, 10] (DFT) and Green's Function (GW) theory [11]. These approaches are related, however, as GW approaches require a set of starting orbitals, and usually orbitals obtained from DFT are used. Typically, the results from the GW calculations are more reliable than their DFT counterparts, with GW often reaching the level of accuracy needed to make predictions of, e.g., spectroscopic transition energies and electrochemical potentials that are reliable enough for the interpretation of material's properties [12]. Optimising DFT calculations therefore is universally relevant.

In this work, we consider the effects of choices made during the implementation of DFT bandgap calculations, seeking for reliable computational approaches that deliver the correct results for the chosen functional. Different approaches are assessed through the calculation of the bandgap for 340 selected 3D materials.

Applications of DFT require the specification of the density functional, and a wide range of choices are currently available that results in calculations of varying accuracy and reliability [13-15]. Seeking reproducible computational approaches, we consider two representative density functionals: the generalized gradient approximation (GGA) designed by Perdew, Burke and Ernzerhof (PBE) [16], and that as modified to produce the PBE0 hybrid functional [17].

Hybrid fucntionals embed some contribution of Hartree-Fock terms to the exchange energy and required significantly more computational resources when implemented in plane-wave codes such as those typically used to study materials.  Each density functional has its own sets of advantages and inadequacies, with the treatment of self-interaction errors and partial occupancies bringing significant issues [18], with methods such as PBE and PBE0 that do not correct the asymptotic potential underestimating, to differing degrees, bandgaps, spectral transition energies, exciton binding strengths, and spectral assignment [19-23].  This work is not concerned about such issues pertaining to the accuracy of density functionals, only to the ability of computational methodologies to deliver reliably the correct answer for the chosen functional.  Use of hybrid functionals such as PBE0 results in increased bandgaps, which can enhance computational stability, but involves new computational aspects that also need to be optimized.

In DFT calculations, it is common practice to represent core electrons implicitly using pseudopotentials (PP).  This can have a large effect in reducing computational demands.  In addition, for heavy elements with inner-core electrons that experience strong relativistic effects, this procedure can remove the requirement for explicit treatment of spin-orbit interactions.  Also, the exclusion of 1s electrons from explicit representation ensures that the electron density varies smoothly near the nucleus.  In all-electron calculations, the electron-nucleus attraction is described by the standard Coulomb potential.  In PP calculations, empirical potentials are used to describe the interactions between the valence electrons and an unpolarizable system containing the nucleus and the core electrons.  For most atoms, a range of PPs are available, differing by the number of electrons incorporated into the core and in the quality of their representation.  Well-tuned PP can facilitate the accurate prediction of properties of solids [24, 25], while significantly reducing computational costs.  In this work, two PPs are considered: a standard one designed for efficient applications involving GGA-type density functionals such as PBE, and an enhanced one designed to give accurate results compatible with the increased accuracy that can be achieved using either GW methods or else more advanced density functionals.

Related to the choice of PP is a second feature that is important in accurate calculations: the specification of the plane-wave basis-set cutoff energy $E_{cut}$. Rapid changes of the electron density with position demands high momentum and therefore kinetic energy for the electrons, which can only be included using extensive plane-wave basis sets. For each atom, a recommended cutoff kinetic-energy is available for each PP, with the cuffoff used in any calculation being the largest value required by any embodied atom. This recommended cutoff energy is designed to give useful results, but more reproducible results can be obtained if $E_{cut}$ is increased. Recognizing this, the values of $E_{cut}$ recommended for GW PPs are systematically increased compared to those for PBE PPs. In addition, in this work, we consider using either the default value for the PP, or else that increased by 30%; such an increase is typical of specifications often used in high-quality calculations [23, 26].

A final key feature of DFT calculations of materials is the integration that must be performed over the Brillouin zone in the three-dimensional reciprocal-lattice space. Usually, a discrete set of vectors, **k**, is used do so this integration. There are several commonly used methods for selecting these vectors, which are known as "*k*-points". The method introduced by Baldereschi [27], extended by Chadi and Cohen [28], and further developed by Monkhorst and Pack [29], is the most widely used approach today, and we exclusively use this approach using Γ-centered meshes. In this approach, the crystallographic directions **a**, **b**, and **c** in reciprocal space are equally divided into $N_a$, $N_b$, and $N_c$ regions, producing $N_a \times N_b \times N_c$ *k*-points at which the DFT wavefunctions need to be evaluated. The accuracy of the calculations of the total energy of the material can be systematically improved by increasing the number of *k*-points used in the integration. From a practical perspective, however, the computational cost is usually linearly related to the number of irreducible *k*-points involved.

A method in common use for the automatic generation of *k*-point grids for materials is that developed by *Wisesa et al.* [30]. In this, a maximum allowable separation in **k**-space between any two points is prescribed; this is known as the *k*-spacing, $k_{sp}$, from which the *k*-point grid is determined using

$$N_i = max(1, ceiling(|b_i|2\pi/k_{sp})).$$

In this equation, *i* represents one of the 3 crystallographic directions, $|b_i|$ is the length of the lattice vector in reciprocal space, and *ceiling(x)* is the ceiling function which returns the least integer that is equal or larger than *x*. In previous studies, $k_{sp}$ = 0.5 Å$^{-1}$ was determined to be a reasonable setting for most calculations [31], and this is currently the default value used by VASP. Nevertheless, setting $k_{sp}$ too large can lead to significant errors, and, in this work, we consider calculations performed using values of $k_{sp}$ = 0.375 Å$^{-1}$, as has been considered appropriate for crystal-structure geometry optimizations [32], and $k_{sp}$ = 0.22 Å$^{-1}$, as has been considered appropriate, e.g., for GW calculations [33]. Nominally, computational cost scales as $k_{sp}^3$, but symmetry can minimize impact.

All observable properties of the material such as the total energy, atomic forces, and spectroscopic transition energies arise through the consideration of all *k*-points used, but useful approximations to electronic-spectroscopic transition energies and electrochemical potentials can be obtained by considering only the orbital energies calculated at each *k*-point. The energy difference $\Delta E$

$$\Delta E = E_{LUMO}(\mathbf{k}_{LUMO}) - E_{HOMO}(\mathbf{k}_{HOMO})$$

between the energy $E_{HOMO}$ of the highest-occupied (HOMO) band maximum $\mathbf{k}_{HOMO}$ and that $E_{LUMO}$ of the lowest-unoccupied band minimum $\mathbf{k}_{LUMO}$ is known as the *orbital bandgap* and provides a crude approximation to the lowest spectroscopic electronic transition energy as well as the difference between the energies needed to add or remove electronic charge. As determination of the orbital bandgap requires knowledge of the HOMO and LUMO orbitals, properties of the entire **k** space are relevant as the occupancies of *k*-point are globally determined. For semiconductors, the HOMO and LUMO are typically well-defined however, making $\Delta E$ simply a function of only $\mathbf{k}_{HOMO}$ and $\mathbf{k}_{LUMO}$. Alternatively, for metals, the orbital-band occupancies are difficult to determine reliably, and to induce stability the parameter $k_\beta T_E$ is usually introduced into the calculations, where $T_E$ specifies a thermal temperature for the electrons and $k_\beta$ is Boltzmann's constant [34]. Variations in the specification of $k_\beta T_E$ can modulate the nature of $\mathbf{k}_{HOMO}$ and $\mathbf{k}_{LUMO}$ and hence modulate the calculated bandgap. If $\mathbf{k}_{HOMO} = \mathbf{k}_{LUMO}$ then only one vector actually needs to be found, and the optical transition is refereed to as being *direct*; otherwise the transition is called *indirect*.

As each *k*-point represents a different symmetry, all *k*-point calculations are, in principle, independent of each other and so their properties can be determined in parallel on modern computer architectures. For GGA-type functionals such as PBE, modern computational strategies implement this principle. However, for hybrid functionals such as PBE0, it is common to expand the entire Hartree-Fock exchange energy about a critical vector in reciprocal space obtained by considering all *k*-points, thus embedding an unexpected dependence of the results at each *k*-point on the selection of the other *k*-points [35]. Ignoring this effect, orbital bandgaps can be determined simply by evaluating the orbital energies at the two specific *k*-points specified by $\mathbf{k}_{HOMO}$ and $\mathbf{k}_{LUMO}$.

The primary effect of variation of $k_{sp}$ on the evaluation of bandgaps $\Delta E$ therefore comes from how closely the points on the resulting *k*-point grids approach $\mathbf{k}_{HOMO}$ and $\mathbf{k}_{LUMO}$. This closeness does not systematically improve as $k_{sp}$ is made smaller. For example, if a direct transition occurs at $\mathbf{k} = (\frac{1}{2}, \frac{1}{2}, \frac{1}{2})$, then using a *k*-point mesh of $N_a = N_b = N_c = 2$ will yield a more reliable result than that obtained from a million-times more expensive calculation using $N_a = N_b = N_c = 201$. For the choices of PP and energy cutoff $E_{cut}$, systematic procedures are available for enhancing the computational reliability, but this is not the case for the specification of the **k** integration parameters based upon $k_{sp}$. In this work, we present a scheme for optimizing $N_a$, $N_b$, and $N_c$ that retain the useful features of $k_{sp}$ for the optimization of energies, forces, and other properties, yet is also systematic in its ability to yield improved bandgaps $\Delta E$.

The influence of the computational parameters such as PP, $E_{cut}$, and $k_{sp}$ operate directly at every possible geometrical structure of a material. In addition, these parameters have an indirect effect in that they also modify the forces and hence the optimal structure predicted by each density functional. Optimizing the reliability of DFT calculations therefore also needs to consider the structure-property relationship for the material.

In summary, in this work, the sensitivity of calculated bandgaps is considered with respect to changes in the *k*-point mesh, $E_{cut}$, and PP as a function of materials structure. In total, nine sets of calculations are performed, named **a – i** in Figure 1. These calculations are connected in the figure by arrows or lines, named **(i)** to **(x)**, that highlight single-parameter changes. Changes that are expected to lead to systematic improvements in the calculations are marked with blue arrows, whereas unsystematic changes are marked with brown lines. For up to 340 selected materials evaluated using the PBE and PBE0 density functionals, calculated changes in the bandgap $\Delta E$ are then statistically analyzed, finding correlations of general use, as well as outliers and their origins.

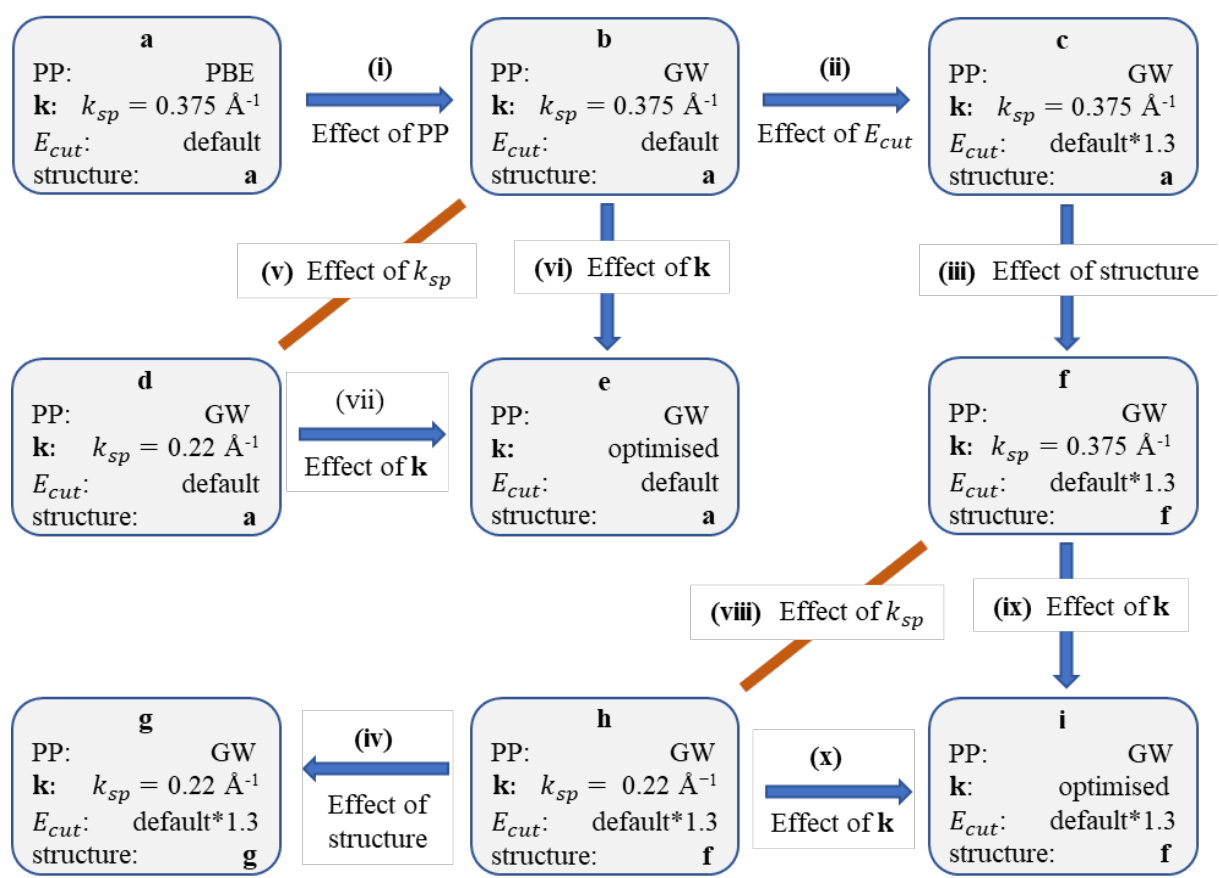

Fig. 1 Overview of the calculations performed, **a – i**, and single-step connections between them, **(i) – (x)**. Connections marked with blue arrows indicate systematic improvements in the methodology, whereas brown connections do not. The structures used are those produced by structural optimizations performed in the stated calculations, either calculations **a**, **f**, or **g**.

# 2 METHODS

## 2.1 Selection of 340 materials.

We started considering the set of 472 3D materials selected by Borlido [36] from within the 2018 Materials-Project (MP) database [37]. Since then, the MP database has been upgraded several times, with some of the 472 selected materials being significantly modified, with poor availability of the original structures. In addition, the modern MP database now lists primitive unit cells instead of the conventional ones accessed by Borlido et al. To avoid structural ambiguities, only structures preserved between the MP 2019 and 2022 databases are considered herin, utilizing the structure as represented in the 2022 MP database. Some of these materials were found to undergo significant structural changes when optimized using PBE, and these were also excluded from consideration. Finally, structures not computationally feasible for reliable bandgap calculations using PBE0 owing to the computational resources required were also excluded. As a result, the number of materials considered was reduced to 340. Even still, not all PBE0 calculations were completed for this set, as detailed in Supplementary Material (SM).

## 2.2 General DFT procedures.

The DFT computations were performed using VASP6.4.3. There are some significant computational parameters other than the three considered herein in detail. The computational algorithm, which in most cases was set to ALGO =ALL, but if this failed then ALGO = NORMAL was used. For PBE0, the results were found to be sensitive to the long-range treatment of the Hartree-Fock exchange component, with reliable results only obtained using the default setting for HFRCUT. Other parameters included PREC = ACCURATE, an electronic loop convergence energy tolerance of $10^{-6}$ eV, and a geometry-optimization convergence criterion requiring the forces on each atom to be less than $10^{-3}$ eV Å$^{-1}$. The geometry optimizations were performed in 4 successive steps, optimizing in term all coordinates with the unit cell (ISIF = 2), the unit cell (ISIF = 6), all properties (ISIF = 3), and finally all atomic coordinates (ISIF = 2) again. All reported properties are obtained using subsequent single-point energy calculations. The wavefunction files were deleted between each computational step to remove memory effects.

## 2.3 A scheme to optimize the choice of *k*-point grids for bandgap calculations.

To find a more robust method for determining $\mathbf{k}_{HOMO}$ and $\mathbf{k}_{LUMO}$, we utlise the computational efficiency of the PBE functional to perform calculations using a 16 × 16 × 16 *k*-points grid for each material. This grid-representation of the band energies was then interpolated to obtain the desired stationary points $\mathbf{k}_{HOMO}$ and $\mathbf{k}_{LUMO}$. For both the HOMO and LUMO, the Hessian matrices $\mathbf{H} = \frac{\partial E}{\partial \mathbf{k}^2}$ were determined numerically at the extremum grid points, as well as evaluating the associated first derivatives $\frac{\partial E}{\partial \mathbf{k}}$. The values of $\mathbf{k}_{HOMO}$ and $\mathbf{k}_{LUMO}$ were then obtained by interpolation, evaluating the differences $\Delta \mathbf{k}$ of the stationary points from the extremum grid points as

$$\Delta \mathbf{k} = -\mathbf{H}^{-1} \cdot \frac{\partial E}{\partial \mathbf{k}}.$$

Given $\mathbf{k}_{HOMO}$ and $\mathbf{k}_{LUMO}$, the task then becomes the determination of a *k*-point mesh that has grid points both sufficiently close to the desired vectors and extensive enough to allow related properties like spectroscopic transition energies to be evaluated. Such properties could be evaluated in subsequent works using e.g., ΔSCF, Green's function, or Bethe-Salpeter approaches. It is often assumed that *k*-point meshes produced by $k_{sp} = 0.375$ Å$^{-1}$ are sufficient for such purposes [12], but smaller values could be used if necessary. What is required then is a *k*-point mesh that is at least as large as the one indicated using $k_{sp}$ yet has grid points close to both $\mathbf{k}_{HOMO}$ and $\mathbf{k}_{LUMO}$.

We consider every possible *k*-point mesh from that generated using $k_{sp} = 0.375$ Å$^{-1}$ up to 16 × 16 × 16. For each possibility, the energy in the HOMO and LUMO band-extremum energies are estimated from the previously determined values of $\mathbf{k}_{HOMO}$ and $\mathbf{k}_{LUMO}$ and the Hessian matrix using

$$error = \frac{1}{2} \Delta \mathbf{k} \cdot \mathbf{H} \cdot \Delta \mathbf{k}$$

where now $\Delta \mathbf{k} = \mathbf{k}_{HOMO \text{ or } LUMO} - \mathbf{k}$ are the deviations from a grid point to the extremum values. If the estimated error in the bandgap is less than some threshold level $\Delta E_T$ which we choose to be 0.025 eV, then that *k*-points mesh is considered as a possible best choice. The best choice is obtained by determining which of the available options has the smallest *k*-points mesh in terms of the product $N_a \times N_b \times N_c$.

# 3 RESULTS AND DISCUSSION

Full details of the results and their processing are provided in SM. This includes ASCII files listing the optimized coordinates of each material by each method plus some critical calculation parameters and results. Also included are Excel files listing 43 parameters or results for each of the 5526 completed calculations, along with their mathematical treatment that generates the results tables and figures.

Statistical summaries of the results from the 10 comparisons indicated in Fig. 1 are listed in Table I. This includes the average deviations between the results (AVE), the mean-absolute deviations (MAD), the standard deviations (STDEV), and the minimum (MIN) and maximum (MAX) deviations. Low magnitudes are found for the AVE and MAD errors, with worst-case results of -0.10 eV and 0.15 eV, respectively. This indicates that, on average, computationally efficient schemes can deliver useful results. Nevertheless, the MIN and MAX deviations range from -2.77 eV to 0.95 eV, indicating that method changes can have profound consequences and hence computationally efficient schemes can deliver significantly unreliable results. Subsequent analysis focuses mostly on unreliability, its causes, and the impact that similar effects could have on general DFT calculations. Table II highlights features of the most significant identified outliers.

Table I. Statistical properties of the comparisons **(i)-(x)** shown in Fig. 1, including the average deviation (AVE), the mean absolute deviation (MAD), standard deviation (STDEV), and minimum (MIN) and maximum (MAX) deviations, all in eV, for up to 340 materials.

|  | PBE | | | | | PBE0 | | | | |
| --- | --- | --- | --- | --- | --- | --- | --- | --- | --- | --- |
| comparison | AVE | MAD | STDEV | MIN | MAX | AVE | MAD | STDEV | MIN | MAX |
| **(i)** PP | -0.01 | 0.02 | 0.03 | -0.07 | 0.10 | 0.01 | 0.03 | 0.04 | -0.07 | 0.16 |
| **(ii)** $E_{cut}$ | 0.00 | 0.00 | 0.01 | -0.03 | 0.14 | 0.00 | 0.00 | 0.01 | -0.04 | 0.04 |
| **(iii)** structure[a] | -0.02 | 0.07 | 0.13 | -1.16 | 0.50 | -0.03 | 0.08 | 0.15 | -1.16 | 0.52 |
| **(iv)** structure | 0.00 | 0.01 | 0.02 | -0.16 | 0.13 | | | | | |
| **(v)** $k_{sp}$[b] | -0.04 | 0.08 | 0.13 | -1.26 | 0.70 | -0.07 | 0.09 | 0.19 | -1.75 | 0.95 |
| **(vi)** k [c] | -0.07 | 0.10 | 0.16 | -2.42 | 0.04 | -0.08 | 0.12 | 0.25 | -2.71 | 0.09 |
| **(vii)** k [c] | -0.02 | 0.04 | 0.08 | -1.16 | 0.11 | -0.01 | 0.05 | 0.11 | -1.21 | 0.23 |
| **(viii)** $k_{sp}$[d] | -0.04 | 0.08 | 0.13 | -1.39 | 0.70 | -0.09 | 0.12 | 0.23 | -1.71 | 0.73 |
| **(ix)** k [e] | -0.07 | 0.10 | 0.16 | -2.77 | 0.07 | -0.10 | 0.15 | 0.28 | -2.71 | 0.09 |
| **(x)** k [f] | -0.03 | 0.04 | 0.08 | -1.38 | 0.10 | -0.02 | 0.07 | 0.15 | -1.21 | 0.19 |

a: 205 materials for PBE0.
b: 339 materials for PBE0.
c: 253 materials for PBE0.
d: 178 materials for PBE0.
e: 192 materials for PBE0.
f: 170 materials for PBE0.

Table II. Outliers in the comparisons **(i) - (iv)**, **(vi)**, **(vii)**, and **(ix) – (x)** (see Fig. 1), listing the calculated bandgaps $\Delta E$ (eV) and the changes between them (eV).[a]

| Comp. | material | calc. | $\Delta E$ | calc. | $\Delta E$ | change |
|---|---|---|---|---|---|---|
| PBE | | | | | | |
| **(iii)** | LiF | c | 10.00 | f | 8.84 | -1.16 |
| | NaF | c | 7.11 | f | 6.15 | -0.97 |
| | Mg2F4 | c | 7.52 | f | 6.84 | -0.68 |
| | CaF2 | c | 7.70 | f | 7.16 | -0.54 |
| | SrF2 | c | 7.27 | f | 6.83 | -0.44 |
| | Tl3AsSe3 | c | 1.16 | f | 0.74 | -0.41 |
| | RbF | c | 5.92 | f | 5.52 | -0.40 |
| | Li2I2O6 | c | 3.30 | f | 3.81 | 0.50 |
| **(ix)** | Bi4Cs6I18 | f | 2.32 | i | 2.40 | 0.07 |
| **(x)** | KTaO3 | h | 2.16 | i | 2.26 | 0.10 |
| PBE0 | | | | | | |
| **(iii)** | LiF | c | 13.45 | f | 12.29 | -1.16 |
| | NaF | c | 10.22 | f | 9.31 | -0.91 |
| | Mg2F4 | c | 10.80 | f | 10.17 | -0.62 |
| | CaF2 | c | 10.63 | f | 10.14 | -0.49 |
| | SrF2 | c | 10.21 | f | 9.74 | -0.47 |
| **(vi)** | Os2As4 | b | 2.04 | e | 2.13 | 0.09 |
| **(vii)** | Fe2P4 | d | 2.27 | e | 2.50 | 0.23 |
| | Os2As4 | d | 1.94 | e | 2.13 | 0.18 |
| | KTaO3 | d | 4.32 | e | 4.45 | 0.13 |
| | AuRb | d | 0.78 | e | 0.90 | 0.12 |
| | InP | d | 1.88 | e | 1.98 | 0.10 |
| | LiZnAs | d | 2.10 | e | 2.19 | 0.10 |
| **(ix)** | Os2As4 | f | 2.04 | i | 2.13 | 0.09 |
| **(x)** | Os2As4 | h | 1.94 | i | 2.13 | 0.19 |
| | KTaO3 | h | 4.32 | i | 4.45 | 0.13 |
| | AuRb | h | 0.82 | i | 0.94 | 0.12 |
| | CdI2 | h | 4.05 | i | 4.16 | 0.11 |
| | InP | h | 2.03 | i | 2.12 | 0.10 |
| | Sb2Te3 | h | 1.14 | i | 1.23 | 0.09 |

a: For comparisons **(i) - (iv)**, outliers are shown of magnitude > 0.4 eV, whereas, for comparisons **(vi)**, **(vii)**, **(ix)**, and **(x)**, all outliers are listed that are > 0.09 eV. Comparisons **(v)** and **(viii)** are not listed the embodied changes are random in nature.

## 3.1 pseudo potential

The direct effect of the PP on calculated bandgaps is examined in comparison **(i)**. For it, Figure 2 shows the correlations obtained for both PBE and PBE0 bandgaps calculated at the same geometry using the same $E_{cut}$ and $k_{sp}$, considering the use of PBE versus GW PPs. Mostly, only small changes are found as a consequence of changing the PP, with Table I listing MAD changes of just 0.02 eV for PBE and 0.03 eV for PBE0. The largest changes found were 0.10 eV for PBE and 0.16 eV for PBE0, changes too small to be included in the outliers table, Table II. Materials containing Ba were particularly sensitive to the quality of the PP, as was $Mo_2Sr_2O_8$ and AuCs.

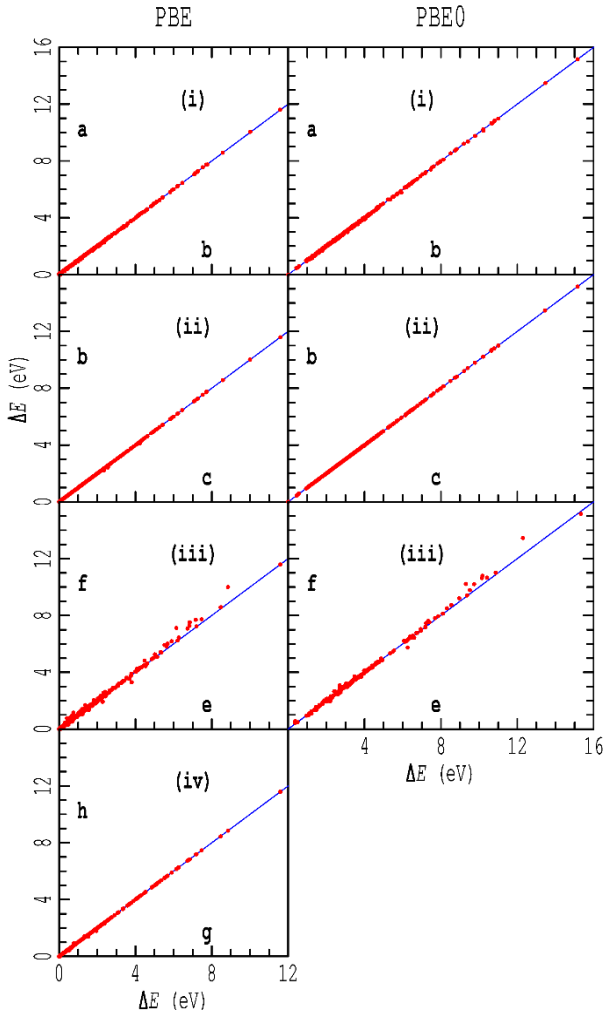

Fig. 2 Comparisons **(i)** to **(iv)** of calculation results **a**, **b**, **c**, **e**, **f**, **g**, and **h** (see Fig. 1), of bandgaps $\Delta E$ obtained using either the PBE or PBE0 density functional.

## 3.2 $E_{cut}$

Increasing the value of the plane-wave basis-set cutoff energy $E_{cut}$ facilitates improved description of the electronic structure in close towards nuclei where electron densities can change quickly. Figure 2 shows the correlations **(ii)** between sets of PBE or PBE0 band gaps calculated using the same PP, structure, and $k_{sp}$, using either the default value for $E_{cut}$ or else this increased by 30 %. Akin to the results found for changing the PP, $E_{cut}$ changes are found mostly have little effect, with MAD changes of just 0.00 eV for both PBE and PBE0. The largest change obtained was 0.14 eV for the PBE calculation of $Bi_4Cs_6I_{18}$, but again this is too small to appear in Table II.

### 3.3 Structural optimization

The comparisons **(iii)** and **(iv)** shown in Fig. 2 depict bandgap changes induced by geometry optimization. It is indeed common practice to use lower-cost approaches to optimize geometries and higher-cost approaches to evaluate properties based upon them, and these comparisons assess the reliability of this methodology. Computational parameters such as PP, $E_{cut}$, and $k_{sp}$ not only have direct effects on bandgaps calculated at the same geometries but also can have indirect effects induced by geometry reoptimization. Specifically, comparison **(iii)** considers the indirect effect of both the PP and $E_{cut}$, whereas **(iv)** considers the computationally expensive effect of changing $k_{sp}$ from 0.375 Å$^{-1}$ to 0.22 Å$^{-1}$ and is considered only using the PBE functional.

Statistical properties such as the MAD changes listed in Table I are small, being 0.07 eV using PBE and 0.08 eV using PBE0 for comparison **(iii)**, with just 0.01 eV using PBE for cmparison **(iv)**. Nevertheless, the MIN and MAX changes can be large, from -1.16 eV to 0.52 eV for **(iii)**, reduced to between -0.16 eV and 0.13 eV for **(iv)**. The most significant outliers are listed in Table II and are often associated with fluorine chemistry, with the use of high quality PPs being required for accurate structural optimizations, as has previously been noted [23]. More broadly, significant errors exceeding the experimentally significant magnitude of 0.1 eV are found, 18% and 21% for PBE and PBE0 calculations for **(iii)** but just 1% of PBE calculations for **(iv)**. In general, the use of crude PPS and low energy cutoffs in structural optimizations appears to be too risky to be used as a standard practice in modern calculations.

## 3.4 Brillouin-zone integration

Figure 3 shows results for each material for comparison sets **(v) – (x)**, obtained using either the PBE or PBE0 functionals. The presentation of the results is modified from that in Figure 2 as now the differences between calculated bandgaps are shown as a function of what would naively be considered to be the poorer calculation. For example, for comparison **(v)**, the abscissa is the bandgap $\Delta E_\mathbf{b}$ from calculation series **b**, whereas the abscissa is the difference in bandgap from these results to those from series **d**, $\Delta E_\mathbf{d} - \Delta E_\mathbf{b}$. Negative differences correspond to the expectation that the better calculation delivers the lower bandgap, and attention focuses on the maximum magnitude of such negative differences, as well as the manifestation of any positive differences.

Comparisons **(v)**, and **(viii)**, marked in orange in Fig. 1, compare calculations performed using $k_{sp}$ = 0.375 Å$^{-1}$ compared to 0.22 Å$^{-1}$. Each calculation interpolates the band energies on a grid, and naively the results from the finer grid are expected to provide a better approximation to the location of the valence and conduction band extrema. Indeed, using $k_{sp}$ = 0.22 Å$^{-1}$ delivers results up to 1.6 eV lower than those from the courser $k_{sp}$ = 0.375 Å$^{-1}$ grid, but the coarser grid could randomly describe the extrema better and is found to produce results that are up to 0.8 eV better. These random results are not highlighted in Table II as they are not considered to be outliers. Nevertheless, they serve to highlight the dangers in using any arbitrary scheme to perform the Brillouin-zone integration. For such an approach to be reliable, the differences found when using different values of the arbitrary parameter (here $k_{sp}$) would need to be small, but the magnitude of the differences between the $k_{sp}$ = 0.375 Å$^{-1}$ compared to 0.22 Å$^{-1}$ results is found to exceed the experimentally relevant bandgap change of 0.1 eV for 16% of materials using PBE and for 23% using PBE0. A higher percentage is expected for PBE0 than for PBE as the band structure is sharper for PBE0 owing to its larger bandgap, and hence inadequacies in **k** have a larger effect. The similarity of the results presented in Figure 3 for comparisons **(v)** and **(viii)** indicate that this effect is independent of other calculation properties, as expected.

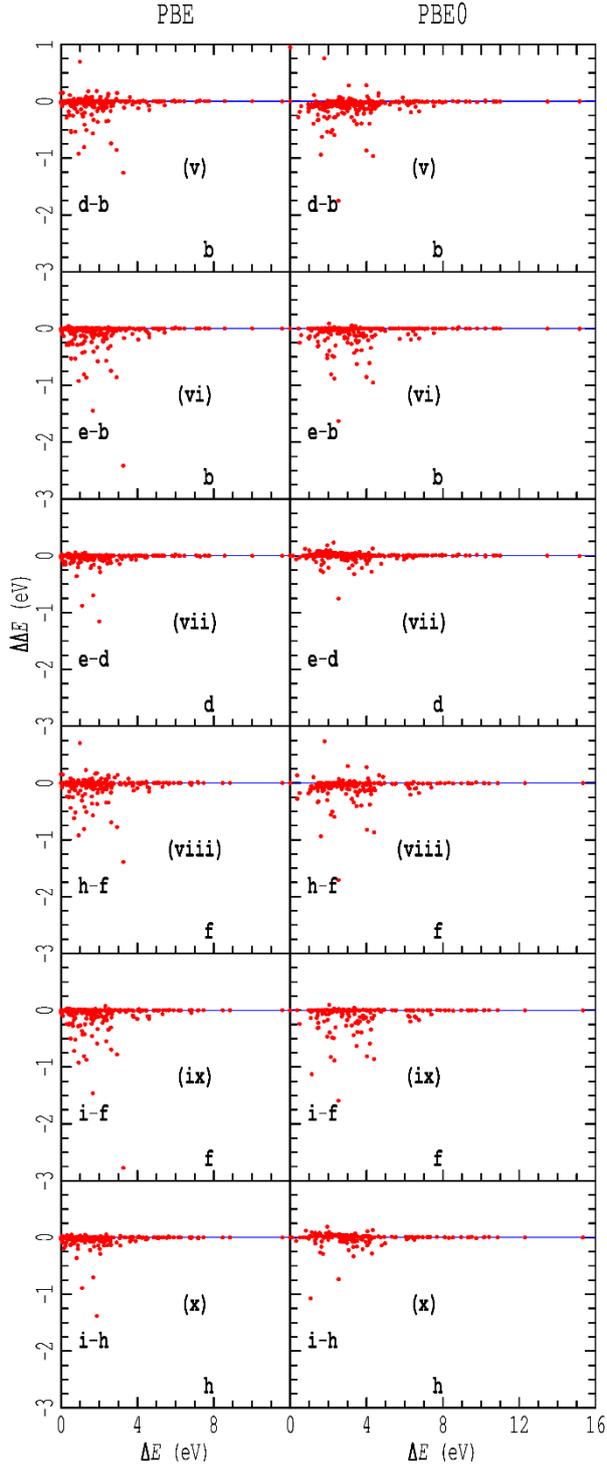

Fig. 3 Distribution diagrams of the bandgap differences ΔΔ$E$ for comparisons **(v)** to **(x)** versus the bandgap Δ$E$ obtained from the more advanced method, involving calculation results **b**, **d**, **e**, **f**, **i**, and **h** (see Fig. 1), obtained using either the PBE or PBE0 density functional.

Comparisons **(vi)**, **(vii)**, **(ix)** and **(x)** compare the results obtained using $k_{sp}$ = 0.375 Å$^{-1}$ or 0.22 Å$^{-1}$ to those obtained using a grid optimized based upon the interpolation of PBE calculations using a 16×16×16 grid. The optimized grid is found to deliver bandgaps up to 2.5 eV lower than those obtained using an arbitrary grid. No results exceeding 0.05 eV, which is twice the error tolerance set in the grid-optimization procedure, are expected, and all results found exceeding 0.09 eV are listed in the outliers table, Table II. Using PBE, only one such outlier is found from 1360 calculations, which could be expected based on random events. This indicates the effectiveness of the interpolation procedure.

Using PBE0, the situation is more complex, however. First, the integration grid that is used was optimized for PBE and then applied to PBE0, so if the band structure has a different form for PBE0 than for PBE, then the interpolation scheme will be inappropriate and significant positive differences could result in Figure 3. Second, the numerical method used by VASP to determine the exchange energy for hybrid functionals involves the selection of a special value of the **k** vector, which is dependent on the entire **k**-point grid used [35]. The general expectation based on symmetry properties that the orbital energies obtained at one **k**-point are independent of any other **k**-point used in the calculation therefore does not hold. Hence this seemingly random effect could induce positive differences (as well as negative ones). For PBE0, only 1.6 % of the comparisons are reported as significant outliers in Table II, all of which can be attributed to one of these two issues. The optimized Brillouin-zone integration technique for hybrid functionals is therefore generally reliable but not flawless.

## 4. CONCLUSIONS

The challenge of optimizing DFT calculation options to return reliable bandgap predictions for a chosen functional was investigated considering four key parameters: the PP, the energy cutoff $E_{cut}$, the details of the geometry of the material, and the Brillouin-zone integration. Based upon analysis of calculations for up to 340 3D materials, the broad conclusion is that commonly applied protocols deliver useful results on average. Nevertheless, for a significant fraction of the calculations, significant shortcomings were identified, questioning the overall reliability of calculations performed using standard computationally efficient protocols.

Firstly, calculations of the bandgap using basic PPs and $E_{cut}$ settings are found to provide excellent approximations for calculations done using the same structure, but when used to perform structural optimizations, these parameters led to significant errors in 18% of cases for PBE and 21% for PBE0. Their continued use in such commonly applied situations is therefore not recommended.

Secondly, automated schemes for optimizing grids using $k_{sp}$ are found to be inadequate for the determination of bandgaps as they suffer from large random errors that are experimentally significant for 16% of the materials studied using PBE and 23% using PBE0. To deliver reliable results, an interpolation scheme based upon PBE calculations for a 16×16×16 grid. This is found to work extremely well for PBE calculations, and to deliver results with only 1.6% of materials manifesting experimentally significant errors when the PBE grid is applied in PBE0 calculations.

**Acknowledgments**

We thank the ARC Centre of Excellence in Quantum Biotechnology, and National Computational Infrastructure (Australia) for provision of computing resources under NCMAS account d63, supported also by the University of Technology Sydney, and ANUMAS account x89.